\pdfoutput=1

\documentclass[final,5p,twocolumn]{elsarticle}

\usepackage{amssymb}
\usepackage{amsmath}
\usepackage{slashed}
\usepackage{hyperref}
\usepackage{microtype}

\usepackage[utf8]{inputenc}

\usepackage{soul,xcolor}

\usepackage{graphicx}
\usepackage{epsfig}
\usepackage{amssymb}
\usepackage{color}
\usepackage{epstopdf}
\usepackage{dcolumn}
\usepackage{relsize}
\usepackage{amsmath}
\usepackage{bm}
\usepackage{dcolumn}
\usepackage{txfonts}
\usepackage{physics}
\usepackage{slashed}
\usepackage{float}
\usepackage[mathscr]{euscript}
\usepackage{lipsum}

\usepackage{graphicx,graphics,epsfig}
\usepackage{amsmath,amssymb,slashed,dcolumn,amsfonts,slashed,
}
\usepackage{comment}
\usepackage{multirow}
\usepackage{mathptmx}      

\def\Xint#1{\mathchoice
   {\XXint\displaystyle\textstyle{#1}}%
   {\XXint\textstyle\scriptstyle{#1}}%
   {\XXint\scriptstyle\scriptscriptstyle{#1}}%
   {\XXint\scriptscriptstyle\scriptscriptstyle{#1}}%
   \!\int}
   
\def\XXint#1#2#3{{\setbox0=\hbox{$#1{#2#3}{\int}$}
     \vcenter{\hbox{$#2#3$}}\kern-.5\wd0}}

\def\dashint{\Xint-}

\biboptions{sort&compress}

\begin{document}

\begin{frontmatter}

\title{Spectral-shift and scattering-equivalent Hamiltonians on a
  coarse momentum grid} \author{Mar{\'\i}a G\'omez-Rocha
}\ead{mgomezrocha@ugr.es}\author{Enrique Ruiz Arriola}\ead{earriola@ugr.es}
\address{Departamento de F\'{\i}sica
  At\'omica, Molecular y Nuclear and Instituto Carlos I de F{\'\i}sica
  Te\'orica y Computacional \\ Universidad de Granada, E-18071
  Granada, Spain.}
\begin{abstract} 
  The solution of the scattering problem based on the
  Lippmann-Schwinger equation requires in many cases a discretization
  of the spectrum in the continuum which does not respect the unitary
  equivalence of the S-matrix on the finite grid. We present a new
  prescription for the calculation of phase shifts based on the shift
  that is produced in the spectrum of a Chebyshev-angle variable.
  This is analogous to the energy shift that is produced in the energy
  levels of a scattering process in a box, when an interaction is
  introduced. Our formulation holds for any momentum grid and
  preserves the unitary equivalence of the scattering problem on the
  finite momentum grid.  We illustrate this procedure numerically
  considering the non-relativistic NN case for $^1S_0$ and $^3S_1$
  channels.  Our spectral shift formula provides much more accurate
  results than the previous ones and turns out to be at least as
  competitive as the standard procedures for calculating phase shifts.
\end{abstract}

\end{frontmatter}



\section{Introduction}

Scattering experiments provide usually the most direct
phenomenological approach to constrain the corresponding dynamics in
hadronic systems. The standard procedure is to determine the
scattering observables, which in the case of rotationally invariant
interactions reduces to the determination of phase-shifts via a
partial wave analysis.  The usefulness of the Hamiltonian method
becomes more evident when dealing with the few-body problem, where one
expects to determine bound states and resonances of multihadron
systems in terms of corresponding potentials.  In practice the
calculation of phase shifts requires in general solving an scattering
integral equation, such as the Lippmann-Schwinger (LS)
equation~\cite{Lippmann:1950zz} in the non relativistic case where for
any given potential $V$ the corresponding S-matrix and hence the phase
shifts are determined.  While this is a valid perspective for theories
where the interaction is known {\it ab initio}, ambiguities arise when
one tries to infer the potential from scattering information, as it is
usually the case in nuclear and hadronic physics. In fact, under an
unitary transformation of the Hamiltonian, the corresponding
phase-shifts remain invariant. This generates a whole equivalence
class of Hamiltonians which are actually compatible with the known
scattering information. The notion of equivalent Hamiltonians was
introduced by Ekstein in 1960~\cite{Ekstein:1960xkd} (see
also \cite{Monahan:1971zc} and \cite{srivastava1975off} for an early
review and \cite{Polyzou:2010eq,Polyzou:2010zz} for implications at
the relativistic level). This is similar to the equivalence under
change of variables in quantum field
theory~\cite{Chisholm:1961tha,Kamefuchi:1961sb}.

Only in few cases, however, can one provide an analytical solution of
the scattering problem, so that one employs often a numerical method
which implies a discretization
procedure~\cite{Noyes:1965ib,Kowalski:1965zz}. In the present work we
will be concerned with the discretization in momentum space, since it
has been widely used in the past and is the only practicable method
for non-local interactions which have been and are commonplace in low
and intermediate energy hadronic physics. This requires the
introduction of a momentum or energy grid, which has a similar effect
to introducing radial boundary
conditions~\cite{Reifman:1955ca,Fukuda:1956zz}, but is in fact a more
general scheme~\cite{DeWitt:1956be} (see
~\cite{Rubtsova:2010zz,Arriola:2014nia,Rubtsova:2015owa,Arriola:2016fkr}
for a modern perspective).  The interest for this kind of methods for
the LS equation started with the work of Haftel and
Tabakin~\cite{Haftel:1970zz} where actually the interest was more in
providing a method to solve the Bethe-Goldstone equation, which is
genuinely non-local, even if the original interaction is local. The
computation of elastic-scattering phase shifts via analytic
continuation of Fredholm determinants, which are isospectral,
constructed using an square integrable basis was introduced in
Ref.~\cite{Reinhardt:1972zz} (for a review see
e.g. \cite{alhaidari2008j} and references therein). A relevant
question is the choice of the particular grid (see
e.g. \cite{eyre1988adaptive} for consideration of adaptive mesh).  We
will take Gauss-Chebyshev grid whose energy eigenvalue problem
corresponds to diagonalizing in a Laguerre
basis~\cite{Heller:1974zz}. We refer to \cite{Deloff:2006hd} for a
comprehensive and self-contained exposition on Chebyshev methods
within the present context. To avoid confusion, ours corresponds to a
{\it radial} one-dimensional momentum grid and {\it not} to a
three-dimensional momentum grid proposed by the so-called KKR
method~\cite{korringa1947calculation,kohn1954solution} in solid state
physics~\cite{jones1985theoretical} and also called the L\"uscher
formula in the relativistic case because of more recent popularity
within lattice-QCD calculations~\cite{Luscher:1985dn,Luscher:1990ux}.

While the physics of a finite momentum grid is that of the bound
states, the continuum limit provides a clear distinction between bound
and scattering states. Actually, in a finite momentum grid important
properties such as the intertwining properties of the Moller wave
operators do not hold~\cite{muga1989stationary}. Moreover, the
momentum grid solution of the LS equation is not invariant under
unitary transformations on the finite grid (see for
instance~\cite{Arriola:2014nia,Arriola:2016fkr}).  The question, of
course, is that since one expects that with a sufficiently fine grid
the continuum limit will be recovered the number of grid points may be
unnecessarily large.

In this letter we provide a method which is in fact rather accurate
for a coarse grid, preserves phase-equivalence and in fact is more
accurate than the standard numerical solution of the scattering
problem  based on the (phase-inequivalent) LS equation.

While the problem we address is fairly general, for illustration
purposes we consider the toy model separable Gaussian potential
discussed previously~\cite{Arriola:2013era,Arriola:2013yca} which
provides a reasonable description of the $NN$ system in the $^1S_0$
and $^3S_1$ partial-wave channels at low-momenta, and supports none or
one (deuteron) bound state, respectively. The extension to coupled
channels (including tensor forces) or resonant and relativistic
systems such as $\pi\pi$ or $\pi N$ scattering requires some
modifications and will be discussed elsewhere (See
~\cite{Gomez-Rocha:2019zkz} for preliminary results).

\section{Scattering on a finite momentum grid}

As already mentioned, scattering occurs in the continuum but numerical
approximation schemes discretize it by a finite momentum grid. We
review here some well known aspects of both formulations in order to
fix our notation.

\subsection{Continuum formulation}

Quite generally we will consider non-relativistic scattering of two
particles with masses $m_1$ and $m_2$ where $H=H_0+V$, $H_0=p^2/2\mu$
and $\mu= m_1 m_2 /(m_1+m_2)$~\cite{landau2008quantum}. Along the
paper we will work in units $\hbar=c=2 \mu=1$ so that the free energy
is given by $E=p^2$. We will assume that the potential is rotationally
invariant so that the total Hilbert space can be decomposed as ${\cal
H}= \oplus_{l=0}^\infty {\cal H}_l$ and work on the partial wave basis
$|p,l,m_l\rangle$~\footnote{To ease the notation we will drop the
angular momentum quantum number $l$ and the third componend $m_l= - l
, \dots , l$ in what follows.}  which is assumed to fulfill the
completeness relation in the Hilbert subspace ${\cal H}_l$
\begin{eqnarray}
{\bf 1} = \frac2{\pi} \int_0^\infty q^2 dq | q \rangle \langle q |  \ .
\end{eqnarray}
Thus,  the action of the 
Hamiltonian on a given state in momentum space is given by 
\begin{eqnarray}
H \psi_l (p)=
p^2 \psi_l (p) + \frac{2}{\pi} \int_0^\infty q^2 dq V_l (p,q) \psi_l (q) \; .
\label{eq:Hpsi}
\end{eqnarray}
where $V_l(p',p)$ are the matrix elements
and the  corresponding Schr\"odinger equation in momentum
space reads,
\begin{eqnarray}
H \psi (p)= E \psi(p) \ ,
\label{eq:schro}
\end{eqnarray}
$E=k^2 \ge 0$ and a bound-state with (negative) eigenvalue $P_\alpha^2 =- B_\alpha$
corresponds to a pole in the scattering amplitude at imaginary
momentum $P_\alpha= i \gamma$.

In the continuum the $S$-matrix is defined as a boundary value
problem for $E \ge 0$
\begin{eqnarray}
  S(E+i \epsilon)= 1 - 2\pi i \delta(E-H_0) T(E+ i \epsilon) \ ,
   \label{eq:Smatrix}
\end{eqnarray}
where we have introduced the 
$T$-matrix which satisfies the scattering equation in operator form,
\begin{eqnarray}
  T(E)= V + V G_0(E) T(E) = V ( 1- G_0 (E) V)^{-1}  \ ,
  \label{Eq:LS}
\end{eqnarray}
where in the second equality we write the exact result. Other
(complex) energy values are defined by analytical continuation.  This
operator satisfies the reflection property $T(E+i
\epsilon)^\dagger = T(E-i \epsilon) $ if $V=V^\dagger$ in
Eq.~(\ref{Eq:LS}) and hence the unitarity condition, $ S(E+
i \epsilon) S(E+i \epsilon)^\dagger = 1 $, follows also from
$V=V^\dagger$ in Eq.~(\ref{eq:Smatrix}). From its definition
$[S,H_0]=0$, and the phase-shift is defined in terms of the
eigenvalues of the S-matrix, so that $S(E) \varphi_\alpha (E) = e^{2
i \delta_\alpha(E)}\varphi_\alpha (E)$ with $H_0 \varphi_\alpha (E) =
E \varphi_\alpha(E)$, where $\langle \varphi_\alpha (E)| \varphi_\beta
(E') \rangle = \delta_{\alpha \beta} \delta (E-E')$. The equivalence
under unitary transformations $U$ follows from the previous equations;
if $[U,H_0]=0$ then $H \to U H U^\dagger $ implies $V \to U V
U^\dagger $ then $T \to U T U^\dagger$ and hence $S \to U S
U^\dagger$, so that $\delta_\alpha$ remains invariant.  For a
rotational invariant interaction, $[S,\vec L]=0$, so that the LS
equation at the partial waves level for $E= k^2/(2\mu)$ reads
\begin{eqnarray}
T_l(p',p)= V_l(p',p)+ \frac{2}{\pi} \int_0^\infty q^2 dq \frac{V_l (p',q)}{k^2-q^2+ i \epsilon} T_l(q,p) \ ,
\label{eq:LS}
\end{eqnarray}
and introducing the reaction matrix via $T = R - i \pi \delta (E-H_0)$
one has 
\begin{eqnarray}
R_l(p',p)= V_l(p',p)+ \frac{2}{\pi} \dashint_0^\infty q^2
dq \frac{V_l (p',q)}{k^2-q^2} R_l(q,p) \ ,
\label{eq:R}
\end{eqnarray}
so that
\begin{eqnarray}
R_l(p,p)= - \frac{\tan \delta_l(p)}{p} \ . 
\label{eq:Rdel}
\end{eqnarray}

\subsection{Discrete formulation}

The previous equations can be solved numerically by restricting the
Hilbert space ${\cal H}_l$ to a finite $N$-dimensional space ${\cal
H}_{l,N}$. On a $N$-dimensional momentum grid, $p_1 < \dots < p_N $,
by implementing a high-momentum ultraviolet (UV) cutoff, $p_{\rm
max}=\Lambda$, and an infrared (IR) momentum cutoff $p_{\rm min}
= \Delta p$ ~\cite{Szpigel:2010bj}. The integration rule becomes
\begin{eqnarray}
\int_{\Delta p}^\Lambda dp f(p) \to \sum_{n=1}^N w_n f(p_n) \, , 
\end{eqnarray}
and the completeness
relation in discretized momentum space reads 
\begin{eqnarray}
1=\frac{2}{\pi}\sum_{n=1}^N w_n p_n^2 | p_n \rangle \langle p_n | \, .
\end{eqnarray}
For instance, the eigenvalue problem on the grid may be formulated as
\begin{eqnarray}
H \varphi_\alpha (p) = P_\alpha^2 \varphi_\alpha (p)   \, ,
\end{eqnarray}
where the matrix representation of the Hamiltonian reads
\begin{eqnarray}
H_{nm} = p_n^2 \delta_{n,m} + \frac{2}{\pi} w_n p_n^2
V_{nm} \, , 
\end{eqnarray}
where $H_{nm} = H(p_m,p_m)$ and $V_{nm} = V(p_m,p_m)$ have been
defined.

In practice we will use the Gauss-Chebyshev quadrature method taking
the corresponding grid
points~\cite{demidovich1973computational,stoer1989numerische}, which
after re-scaling to the interval $[0,\Lambda]$ read,
\begin{eqnarray}
p_n &=& \Lambda \sin^2 \left(\phi_n/2 \right) \label{eq:pn} \ , \\
\phi_n &=& \frac{\pi}N (n-1/2) \ , \\
w_n &=&  { \Lambda \over 2} \sin \phi_n \Delta \phi_n \ , \\
\Delta \phi_n &=& \frac{\pi}{N} \ , \label{eq:dphin}
 \end{eqnarray} where we have introduced the {\it Chebyshev angle},
$\phi_n$, which will play a crucial role in our considerations bellow.
\noindent
From these definitions we have 
\begin{eqnarray}
p_{\rm min} = p_1 &=&  \Lambda \sin^2 \left( \frac{\pi}{4N}\right) \ ,
\;  \\ 
p_{\rm max} = p_N &=&  \Lambda \sin^2 \left[ \frac{\pi}{2N} (N-1/2) \right]  \; .
\end{eqnarray}
\noindent
As it is well known, this grid choice guarantees an exact result for
polynomials in $p$ to order $M \le N$, i.e.
\begin{eqnarray}
\int_{\Delta p}^\Lambda dp \frac{P_M (p)}{\sqrt{\Lambda^2-p^2}}  = \sum_{n=1}^N w_n \frac{P_M(p_n)}{\sqrt{\Lambda^2-p_n^2}} \; .
\end{eqnarray}
\noindent
Taking matrix elements on the momentum grid of the LS equation in operator form we get
\begin{eqnarray}
T_{nm}(p) = V_{nm} + \frac2{\pi} \sum_{k=1}^N w_k \frac{p_k^2}{p^2-p_k^2+ i \epsilon} V_{nk} T_{k,m}(p) \; .
\label{eq:Tnm}
\end{eqnarray}
where $p^2$ is the scattering energy. The on-shell limit is obtained
by taking $p=p_l$ on the grid. As usual we switch to the reaction
matrix which on the grid yields the equation for the half-on-shell amplitude
\begin{eqnarray}
R_{nm}(p_m) = V_{nm} + \frac2{\pi} \sum_{k \neq m} w_k \frac{p_k^2}{p_m^2-p_k^2} V_{nk} R_{k,m}(p_m) \; ,
\end{eqnarray}
where the excluded sum embodies the principal value prescription of
the continuum version in the limit $\epsilon \to 0$. This equation can
be solved by inversion for any grid point $p_n$ and thus we may obtain
the phase-shifts
\begin{eqnarray}
-\frac{\tan \delta^{\rm LS}(p_n)}{p_n} = R_{nn} (p_n) \; ,
\label{eq:ps-LS}
\end{eqnarray}
where the supper-script LS denotes that these phase-shifts are obtained from
the solution of the LS equation on the grid. Of course, the limit $N
\to \infty$ should be understood in the end. As mentioned, 
one drawback of the LS formulation is the fact that if we undertake a
unitary transformation on the grid, $U_{nm}$ of the Hamiltonian the,
the $T_{nm}(p)$ in Eq.~(\ref{eq:Tnm}) still transforms as its
continuum counterpart (for any $\epsilon)$ but the phase-shifts given
by Eq.~(\ref{eq:ps-LS}) {\it do not} remain invariant due to the
principal value prescription~\footnote{This has been explicitly shown
within the context of the similarity renormalization
group~\cite{Arriola:2014nia,Arriola:2016fkr}.}. Hence the LS
phase-shifts are not isospectral.

\section{Spectral shifts: the transition from the discrete to the continuum}

In the previous section we have described how the scattering equations
can be solved by discretizing the spectrum, thus violating the
isospectrality of the phase-shifts.  In this section we discuss three
alternative spectral shifts: the momentum shift, the energy shift and
the Chebyshev-angle shift which enjoy the isospectrality of the phases.

\subsection{Momentum-shift prescription: scattering in a spherical box}
\label{app:Newton}

The momentum-shift prescription proposed by Fukuda and Newton long
ago~\cite{Reifman:1955ca,Fukuda:1956zz} is the simplest and assumes
that the scattering process takes place in configuration space in a
large spherical box of radius $R$. For a potential with a finite range 
$a$  the reduced wave function $u_l(r )$ has the following asymptotic behavior for $R \ge r \gg a$  
\begin{eqnarray}
u_l(r )& \sim & \sin \left( pr - {l \pi\over 2} + \delta_l(p) \right) \ ,
\end{eqnarray} 
and must vanish for $r=R$, so that $u_l(R)=0$, which implies that
\begin{eqnarray}
pR - {l  \pi \over 2} + \delta_l(p) & = & n\pi \ .
\label{eq:npi}
\end{eqnarray}
In a finite momentum grid of $N$ points the equation yields $N$
eigenfunctions, and thus Eq.~(\ref{eq:npi}) holds for every $p_n$. In
absence of interactions, $\delta_l=0$, and one has
\begin{eqnarray}
p_n R - {l \pi \over 2}  & = & n\pi \ ,
\label{eq:pfree}
\end{eqnarray}
with $n=1, \dots, N$, and $\Delta p_n \equiv p_{n+1}-p_n={ \pi \over
R}$ is the separation of the grid points.  Representing now the
distorted momentum by $P_n$, and using Eq.~(\ref{eq:pfree}) to replace
$n \pi + l \pi /2$, one can write
\begin{eqnarray}
P_nR  + \delta_l(P_n) & = & p_n R \ ,
\end{eqnarray}
which, using that $\Delta p_n = \pi/R$ becomes
\begin{eqnarray}
\delta (P_n) & = & - \pi { P_n - p_n \over \Delta p_n}= - \pi { \Delta P_n \over \Delta p_n} \ .
\label{eq:pshiftd}
\end{eqnarray}
This is the \textit{momentum-shift} formula which strictly speaking
holds for an uniformly distributed momentum grid. This is equivalent
to a trapezoidal rule quadrature which is generally a poor integration
method. In Section \ref{sec:num} we will consider
Eq.~(\ref{eq:pshiftd}) for the Chebyshev grid quadrature method, namely
\begin{eqnarray}
\delta^{\rm MS}(P_n) & = & - \pi { P_n - p_n \over w_n} \ .
\label{eq:pshift}
\end{eqnarray}

\subsection{Scattering in an energy-equidistant discretized spectrum}
\label{app:dewitt}

An alternative approach was proposed simultaneously de
DeWitt~\cite{DeWitt:1956be}.  Let us present DeWitt's argument in a
slightly different way so that our points can be easily formulated.
For the sake of clarity and the benefit of the reader we try to be
pedagogical here since we found some parts hard to follow.
Let us consider the eigenvalue problems 
\begin{eqnarray}
  H \psi_n &=& E_n \psi_n \ , \\
  H_0   \psi_n^{(0)} &=& E_n^{(0)} \psi_n^{(0)} º ,
\end{eqnarray}
with $H=H_0+V$. Our notation is such that $E_n \to E_n^{(0)}$ when $V
\to 0$, and as according to the Landau-von Neumann theorem, there is no crossing between non-degenerate levels~\cite{Landau:1991wop}. 

The cumulative number associated to the discretized Hamiltonian $H$
reads
\begin{eqnarray}
N(E) = \sum_n \theta(E-E_n) = {\rm Tr} \theta(E-H) \ ,
\label{eq:ncum}
\end{eqnarray}
where we have introduced the trace ${\rm Tr} A = \sum_n \langle \psi_n
| A | \psi_n \rangle$. For a discrete spectrum this function
represents a staircase, with unit jumps at any eigenenergy, $E_n$.
In order to proceed further it is important to separate the states
into positive and negative energy states. In the continuum limit the
negative energy states will become {\it bound} states whereas the
positive energy states will become {\it scattering} states. Starting
from the free Hamiltonian where all energies are positive,
$E_n^{(0)}>0$, with a gradually increasing attractive interaction some
energy levels may drive into the negative energy spectrum. From a
variational point of view the discretized energy provides an upper
bound of the true spectrum, since the discretization procedure may be
viewed as a restriction on the physical Hilbert space, $E_n^N \ge
E_n^\infty$ and thus the net effect of the finite grid is {\it
  repulsive}.  In what follows we will assume that the grid is fine
enough so that no positive energy level will cross zero.

The step function in Eq.~(\ref{eq:ncum}) can be regularized as
proposed in Ref.~\cite{RuizArriola:1993tk,Caro:1994ht}, namely
introducing a small imaginary energy $i \epsilon \to i 0^+$, as follows 
\begin{eqnarray}
\frac1{\pi} \Im \log (-x+i\epsilon) &=& \frac12 + \frac1{\pi}\tan^{-1}
(x/\epsilon) \nonumber \\ &\to& \theta(x) \ ,
\end{eqnarray}
where for a general complex number $z = \rho e^{i \theta}$ with $-\pi
\leq \theta \le \pi$ we define the logarithm
\begin{eqnarray}
  \log z = \log \rho + i \theta  \ ,
\end{eqnarray}
where the branch cut runs along the negative real axis. Deriving with
respect to $x$ we also get
\begin{eqnarray}
  \frac1{\pi} \Im \frac1{x-i\epsilon} &=&  \frac1{\pi} \frac{\epsilon}{x^2+ \epsilon^2}
\nonumber \\ &\to& \delta(x) \ ,
\end{eqnarray}
where it is important in what follows to keep a finite $\epsilon$ and to
take the continuum limit $N \to \infty$ with
$\Delta e_n \equiv E_{n+1}^{(0)}-E_{n}^{(0)} \to 0$. 

Then we have,
the regularized cumulative number
\begin{eqnarray}
N(E- i \epsilon) &=& \frac1{\pi} \sum_{n=1}^N \Im \log ( E_n - E + i \epsilon ) \\ 
&=& \frac1{\pi} \Im \Tr \log (H-E + i \epsilon)  \\
&=& 
\frac1{\pi} \Im \log {\rm Det} (H-E + i \epsilon)
\end{eqnarray}
where the identity $\log {\rm Det} A = {\rm Tr} \log A$ has been used.
Using a similar equation for $H_0$ and its corresponding cumulative
number $N_0$ we get for the difference,
\begin{eqnarray}
\Delta N (E- i \epsilon) & \equiv & N(E- i \epsilon)-N_0(E- i \epsilon)\nonumber \\  &=& 
\frac1{\pi} \Im \log {\rm Det} \left[(H-E + i \epsilon) \right. \nonumber \\
 && \left. \qquad \times (H_0-E+i\epsilon)^{-1} \right] \nonumber \\
&=& \frac1{\pi} \Im \log {\rm Det} \left[1- G_0 (E-i \epsilon) V \right] \nonumber \\
&=& \frac1{\pi} \Im \log {\rm Det} \left[1- V G_0 (E-i \epsilon)  \right] \ ,\ ,
\label{eq:DNlog}
\end{eqnarray}
where $G_0 (E - i \epsilon )= (E-i \epsilon -H_0)^{-1}$ is the
resolvent of the free Hamiltonian and the cyclic property of the trace
can be used to allocate $V$ to the left or to the right of $G_0$. The
discontinuity of this function is defined as
\begin{eqnarray}
  {\rm Disc} \, \Delta N &\equiv& \Delta N(E- i \epsilon)-
  \Delta N(E+ i \epsilon) \ .
\end{eqnarray}
Direct application of the eigenvalues yields, 
\begin{eqnarray}
  {\rm Disc} \Delta N =-
\frac2{\pi} \sum_n  \left\{\tan^{-1} \left[ \frac{E_n-E}{\epsilon} \right]
  -\tan^{-1} \left[ \frac{E_n^{(0)}-E}{\epsilon} \right] \right\}
\end{eqnarray}
In order to carry out the sum we use the trigonometric identity,
\begin{eqnarray}
 \tan^{-1} (x)-\tan^{-1} (y) = \tan^{-1} \left[ \frac{x-y}{1+xy}\right] \ ,
\end{eqnarray}
so that  we get 
\begin{eqnarray}
  {\rm Disc}  \Delta N = \frac2{\pi}\sum_{n} \tan^{-1} \left[ \frac{
      (E_n^{(0)}-E_n)/\epsilon}{1+(E_n^{(0)}-E)(E_n-E)/\epsilon^2 }
    \right] \ .
\end{eqnarray}
In this formula the continuum limit $ \Delta e \to 0$ has to be taken
  {\it before} the limit $\epsilon \to 0$. In the limit $ \Delta e
  / \epsilon
\ll 1 $ we change the summation into an integral, $\sum_n \to \int
dn$.  One crucial aspect in DeWitt's formulation is the explicit use
of a uniform energy spectrum, so that one takes $ E_n^{(0)}= n \Delta
e$ and thus $E_n = n \Delta e + \Delta E_n $ where $\Delta E_n$ is the
{\it energy shift}. Clearly, in the continuum limit $\Delta e \to 0$
the energy shift vanishes $\Delta E_n$, but the ratio $\Delta E_n /\Delta e$
remains finite. Defining the change of variables $t= n \Delta
e/\epsilon $, in the limit $\Delta E_n/\epsilon \to 0$, one gets, after
shifting the integration variable,
\begin{eqnarray}
  {\rm Disc}\,  \Delta N &=& -\frac2{\pi} \frac{\epsilon}{\Delta e}\int_{-\infty}^\infty dt
  \tan^{-1} \left[ \frac{ \Delta E /\epsilon}{1+t^2+t \Delta E /\epsilon} \right] \\
  &\to &
  -\frac2{\pi} \frac{\epsilon}{\Delta e}\int_{-\infty}^\infty dt
   \left[ \frac{ \Delta E /\epsilon}{1+t^2} \right] \\
   &=& -2 \pi \frac{\Delta E}{ \Delta e} \ . 
   \label{eq:dN}
\end{eqnarray}
In order to connect this sum with scattering information we 
take into account Eq.~(\ref{eq:DNlog}) and
note that the r.h.s of
Eq.~(\ref{eq:Smatrix}) can then be written as
\begin{eqnarray}
&&  (1- G_0(E-i\epsilon) V)(1-G_0 (E+i \epsilon) V)^{-1}  \nonumber \\
  &=&  1 - [ G_0(E-i\epsilon)- G_0(E+i\epsilon) ] V (1-G_0 (E+i \epsilon) V)^{-1}
\nonumber \\  &=&  1-
2 \pi i \delta(E-H_0) T(E+ i \epsilon)  = S(E+ i \epsilon) \, , 
\end{eqnarray}
 so that using $G_0(E-i\epsilon)- G_0(E+i\epsilon) = 2 \pi
 i \delta(E-H_0)$ we get
\begin{eqnarray}
  {\rm Disc} \, \Delta N = \Im \log {\rm Det} S(E+i \epsilon) = 2 \delta(E) \ ,
\end{eqnarray}
where we have used that the S-matrix is energy diagonal in the
continuum, i.e. $S(E) \varphi_\alpha (E')=0$ for $E' \neq E$.  Merging
this result and the previous Eq.(\ref{eq:dN}) we finally get
\begin{eqnarray}
  \delta (E)&=& - \pi \frac{\Delta E}{ \Delta e} \ , 
  \label{eq:DeltaE}
\end{eqnarray}
where the energy dependence is in the energy shift.  We stress that
this formula holds for an equidistant energy grid, which would
correspond to a trapezoidal rule in energy. If we write it in terms
of momentum variables for the purpose of applying the Gauss-Chebyschev
grid defined by Eqs.~(\ref{eq:pn})-(\ref{eq:dphin}) we get
\begin{eqnarray}
\delta_n^{\rm ES} = - \pi \frac{P_n^2- p_n^2}{2 w_n p_n} \ .
\label{eq:dewitt}
\end{eqnarray}

\subsection{Chebyshev-shift}

In order to profit from {\it both} the use of the Gauss-Chebyschev grid
{\it and} the previous DeWitt's formula, we introduce the angle
$\phi$ given by 
\begin{eqnarray}
 p= \frac{\Lambda}2 (1- \cos \phi) \ , \qquad 0 \le \phi \le \pi  \ ,
\end{eqnarray}
so that the Gauss-Chebyschev grid provides an equidistant angle. Thus,
if we use $\phi$ as the independent variable we may apply DeWitt's
argument {\it mutatis mutandis} for a cumulative number
$N(\phi)$. Using the analogy between the energy levels of scattering
states in a box and the discretization given by a finite grid, and
observing that the equidistance happens in the argument of the cosinus
function, we prescribe the following formula based on the shift of
such an angle, and write:
\begin{eqnarray}
\delta_n = - \pi \frac{\Phi_n- \phi_n}{\Delta \phi_n} = - \pi \frac{\Delta \Phi_n}{\Delta \phi_n} 
\ . 
\end{eqnarray}
where $\phi_n={\pi\over N}\left(n - {1\over 2}\right)$,
$d\phi_n={\pi\over N}$, and the ``distorted'' angles $\Phi_n$ are
calculated from Eqs.~(\ref{eq:pn})-(\ref{eq:dphin}) replacing $p_n$ by
$P_n$. Thus,
\begin{eqnarray}
\delta_n^{\Phi S} =  \frac{2N}{\Lambda} \left[\sqrt{p_n(\Lambda-P_n)}-\sqrt{P_n(\Lambda-p_n)}
\right]
\label{eq:phishift}
\end{eqnarray}
This is the main result of this paper.

One might think that using a similar change of variables in the LS
equation might alter its convergence properties, since discretization
and reparametrization are generally non commutative operations.  Note
however that, from Eq.~(\ref{eq:pn}) one has
\begin{eqnarray}
dp =   \frac{\Lambda}2 \sin \phi d \phi \ .
\end{eqnarray}
On the Chebyshev grid one has $dp_n = w_n$ if $d \phi_n = \pi/N$, so
that this change of variables {\it does not} modify the original
discretized equation in momentum space.

\subsection{Bound state  modifications}

The occurrence of a bound state modifies the formulas in the case of the
energy shift, since direct application of the differences violates the
well known Levinson's theorem~\cite{Ma:2006zzc}, which in the continuum
becomes
\begin{eqnarray}
\delta_l (0) - \delta_l (\infty) = n_l \pi \ , 
\end{eqnarray}
with $n_l$ the number of bound states. In the discrete case,
there appears a discrete momentum scale which requires some re-ordering
of the states~\cite{Arriola:2014nia,Arriola:2016fkr}. In the case of the
energy shift it becomes 
\begin{eqnarray}
\delta^{\rm ES}(p_n) =
\begin{cases}
- \pi \frac{P_{n+1}^2-p_n^2}{2 w_n p_n} \quad {\rm if}\; \quad n < n_{\rm BS}  \\ \\
- \pi \frac{\bar P_{n_{\rm BS}}^2-p_n^2}{2 w_n p_n} \; \quad {\rm if} \quad n=n_{\rm BS} \\ \\
- \pi \frac{P_n^2-p_n^2}{2 w_n p_n} \qquad {\rm if}\; \quad n > n_{\rm BS}
\end{cases}
\label{deltaweg}
\end{eqnarray}
\noindent
where $\bar P_{n_{\rm BS}}^2=(P_{n_{\rm BS}+1}^2+P_{n_{\rm
BS}-1}^2)/2$. Note that in this prescription only the eigenvalues
$P_{n}^2$ corresponding to momenta $p_n < p_{n_{\rm BS}}$ are shifted
one position to the left. The cases for the p-shift and $\phi-$shift
are similar.

\begin{figure*}
\includegraphics[scale=0.45]{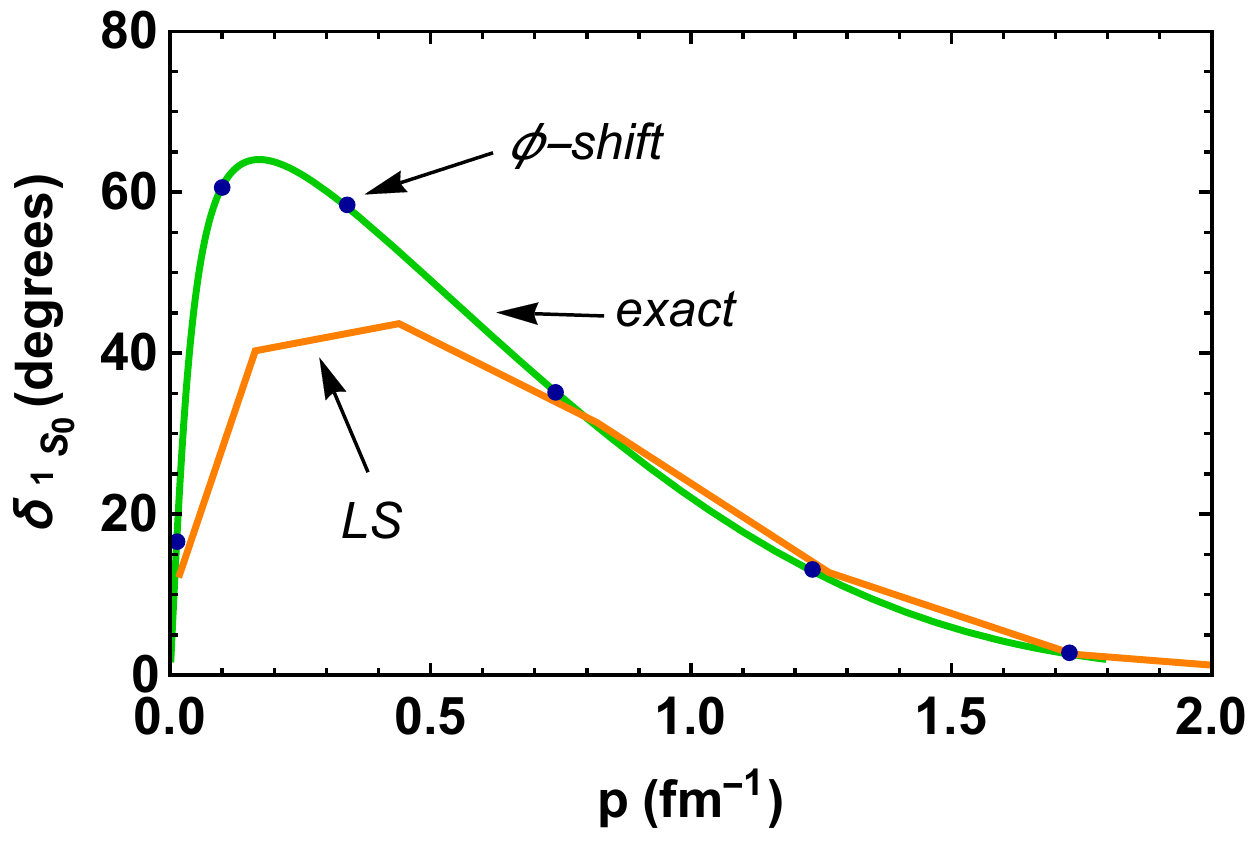}
\includegraphics[scale=0.45]{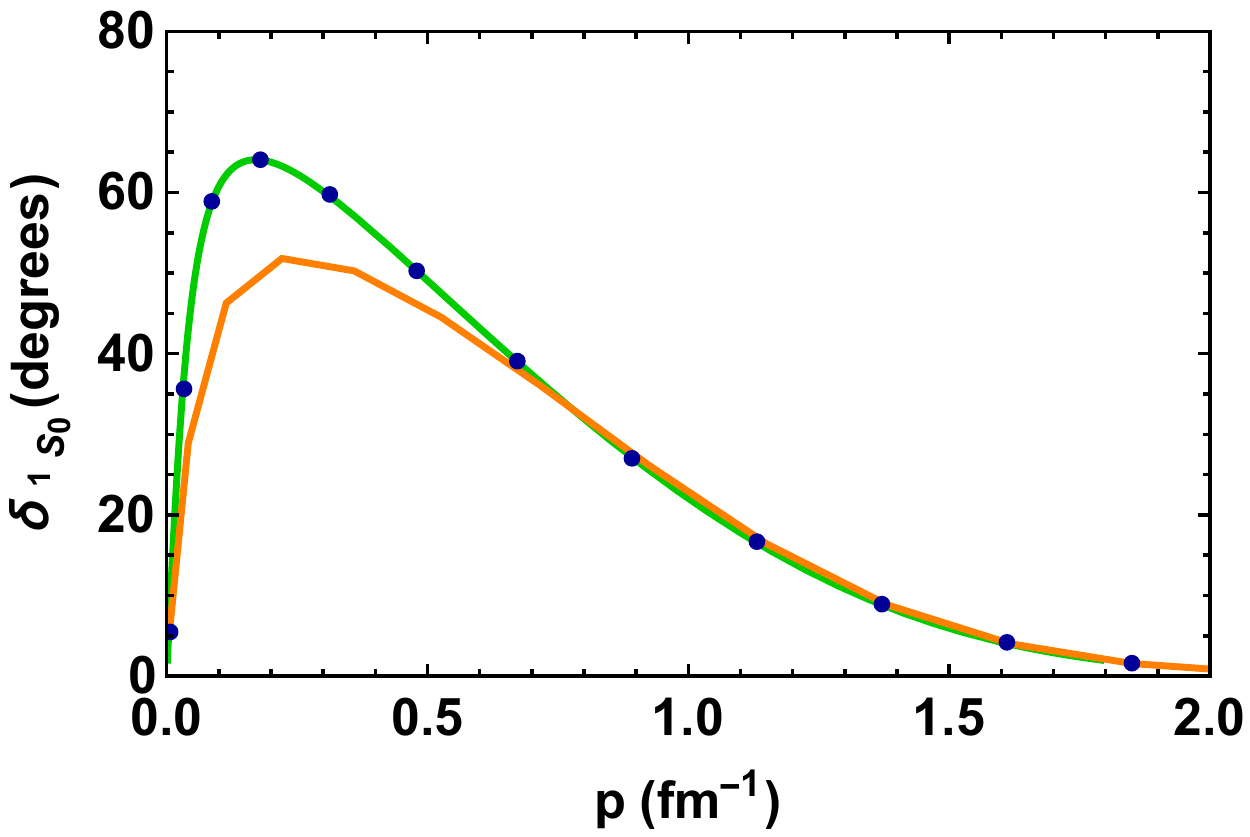}
\includegraphics[scale=0.45]{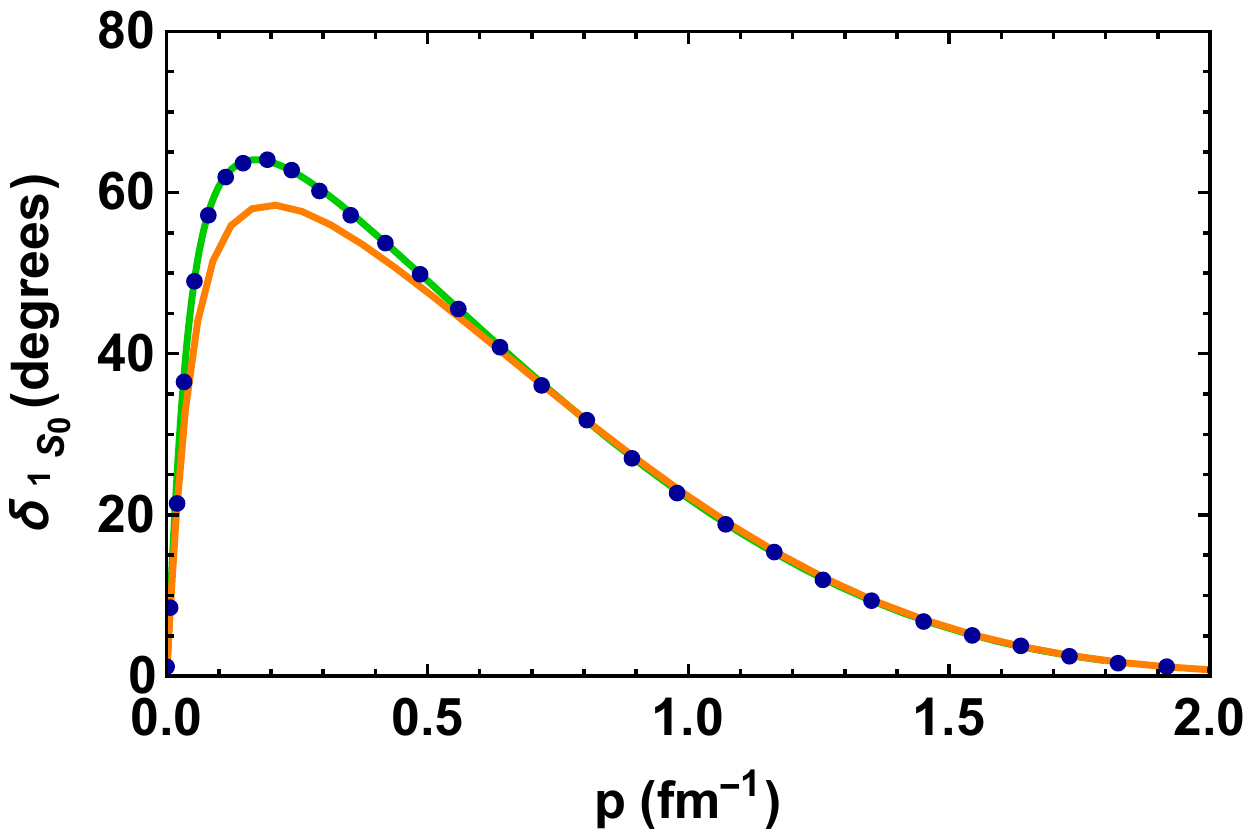}

\includegraphics[scale=0.45]{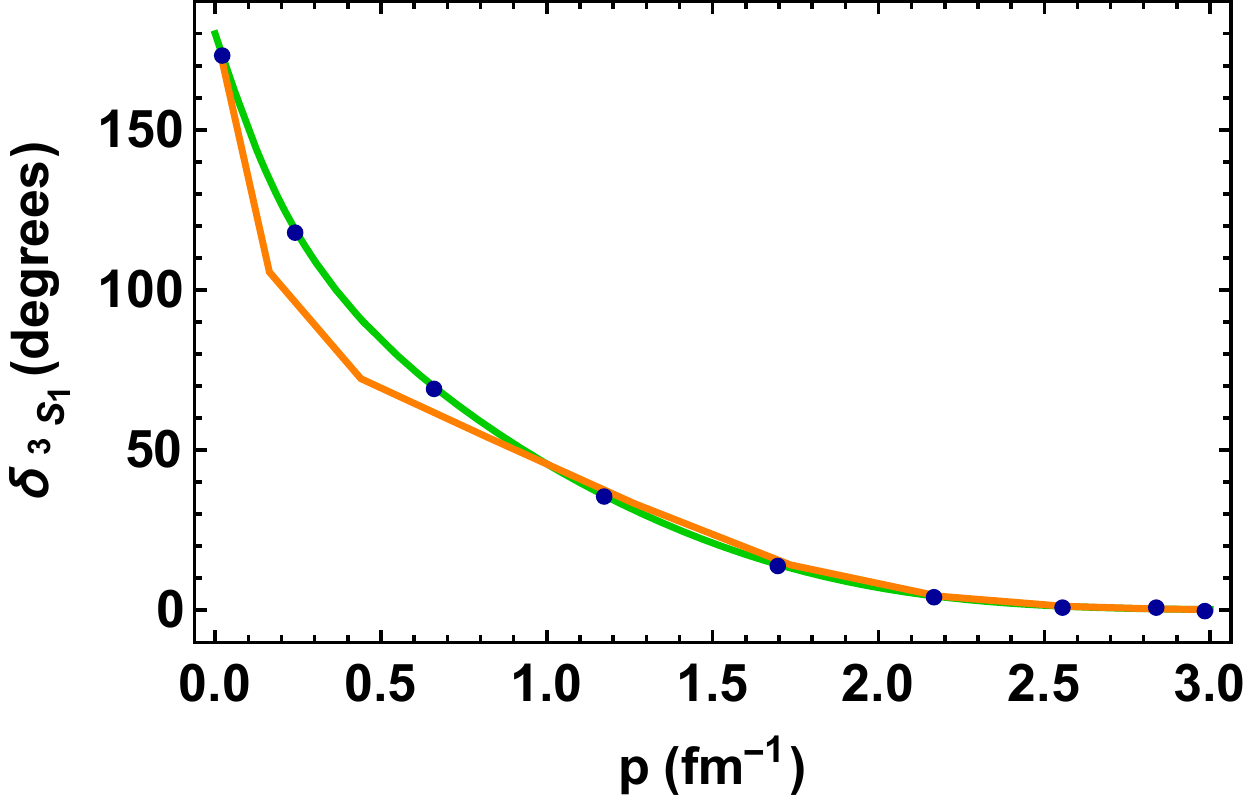}
\includegraphics[scale=0.45]{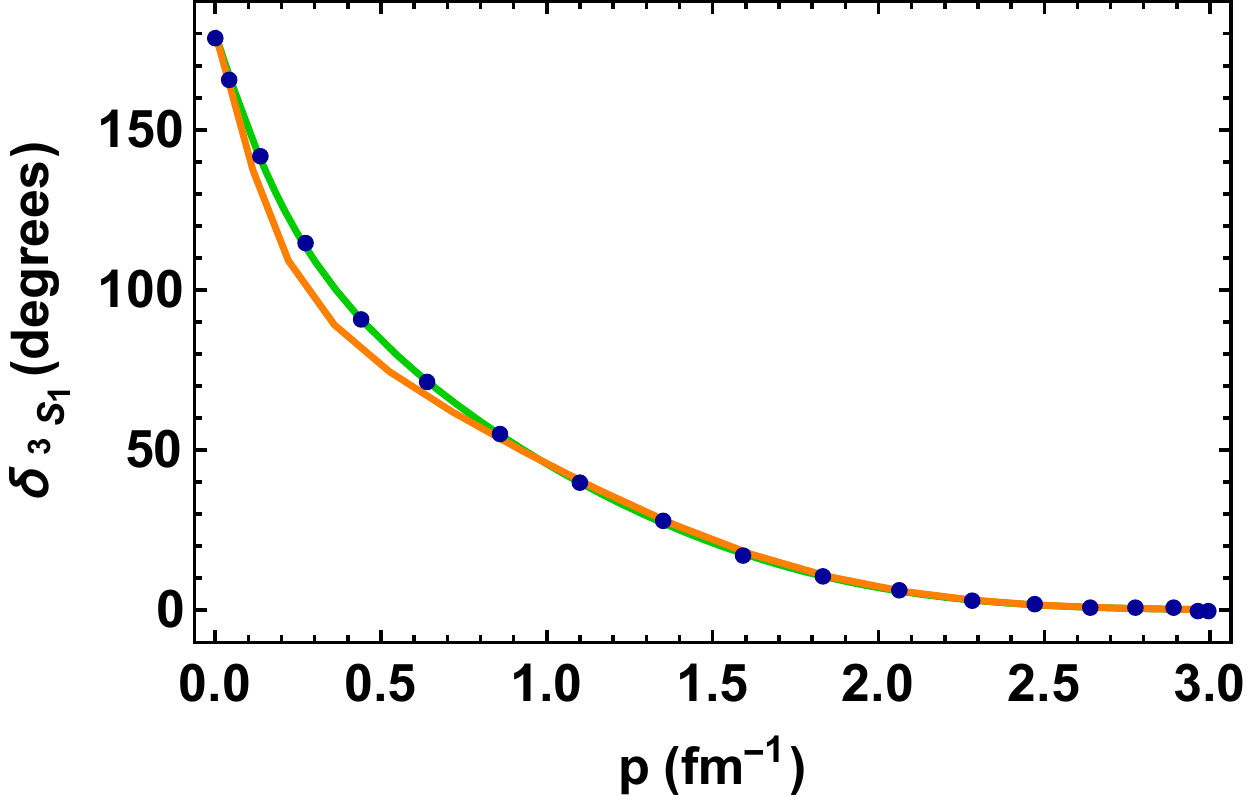}
\includegraphics[scale=0.45]{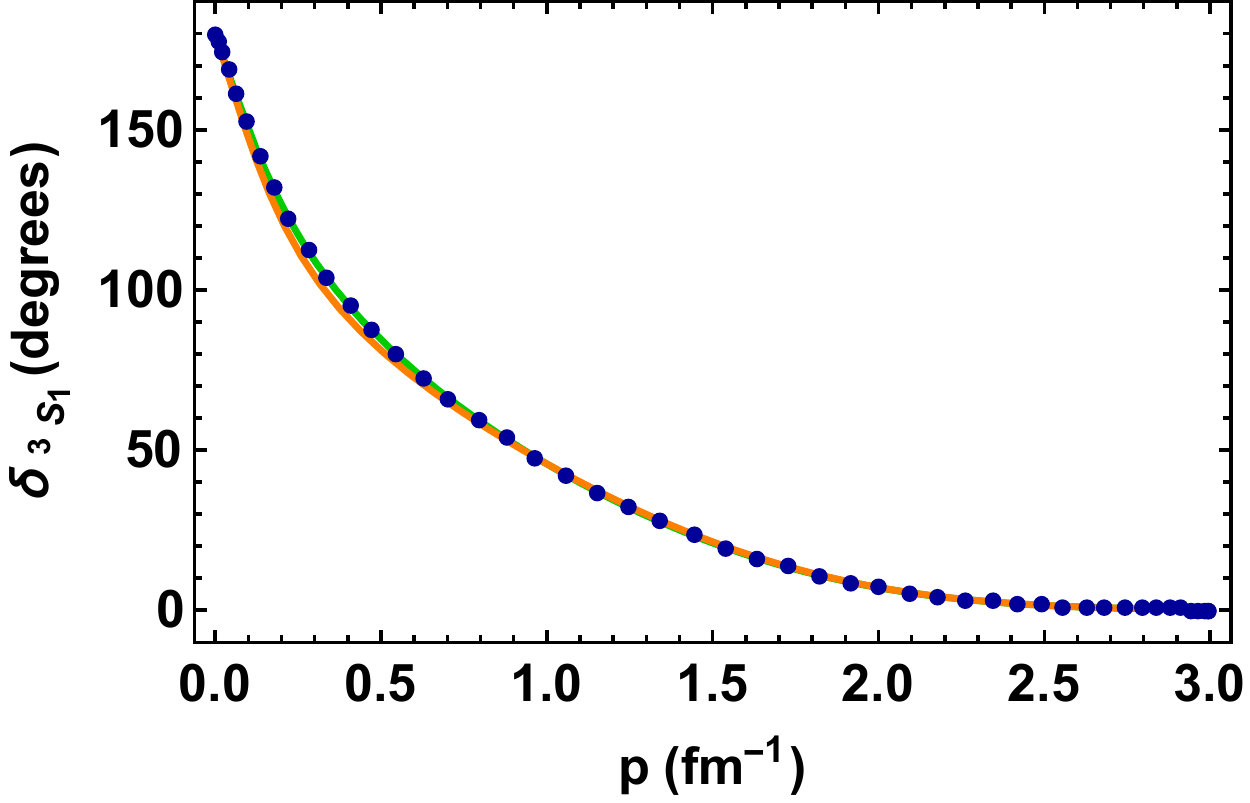}

\caption{(Color online) Comparison of results using our $\phi$-shift prescription (blue dots) with the numerical fit (green, smooth line) and with the result obtained solving the LS equation (orange). Each column corresponds to the same calculation using a grid of $N=$10, 20, and 50 points, respectively. }
\label{Fig:comparisons}
\end{figure*}

\begin{figure*}
\includegraphics[scale=0.45]{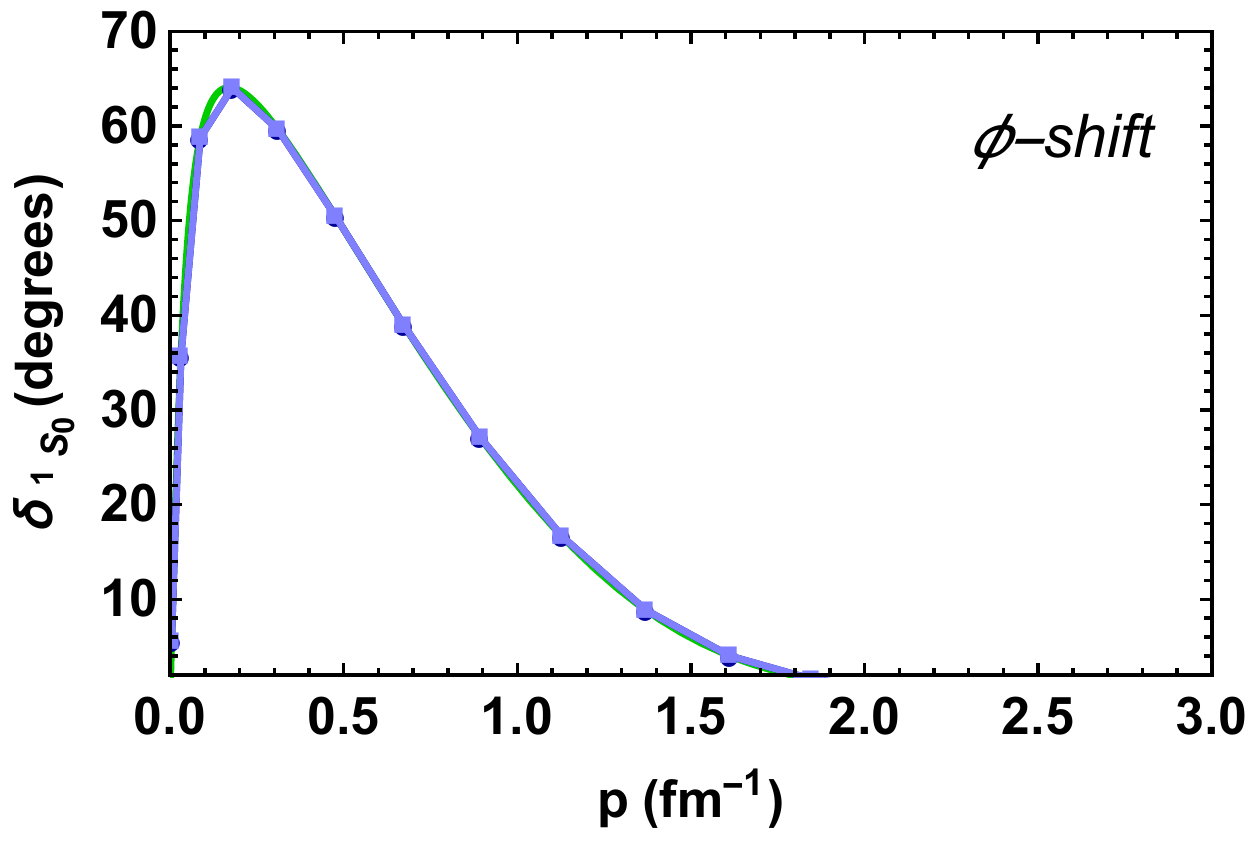}
\includegraphics[scale=0.46]{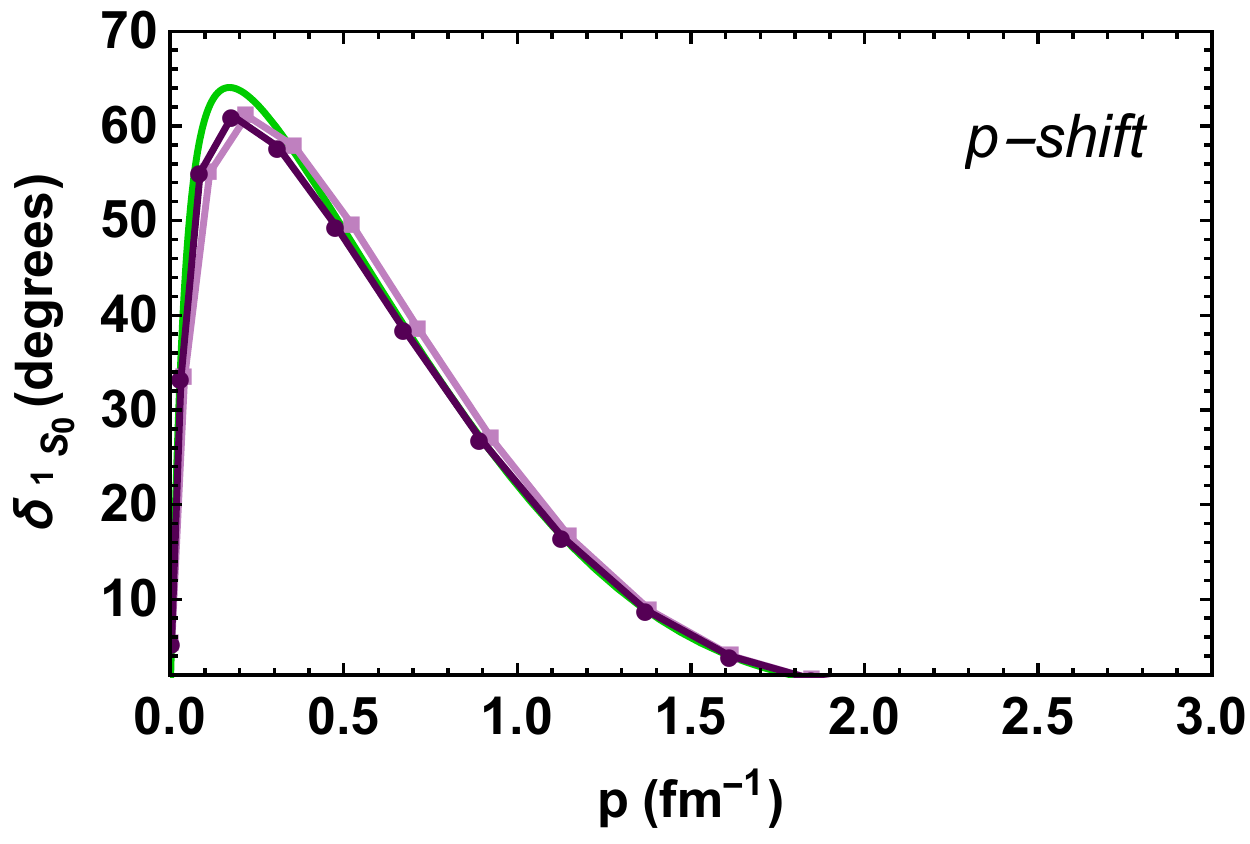}
\includegraphics[scale=0.46]{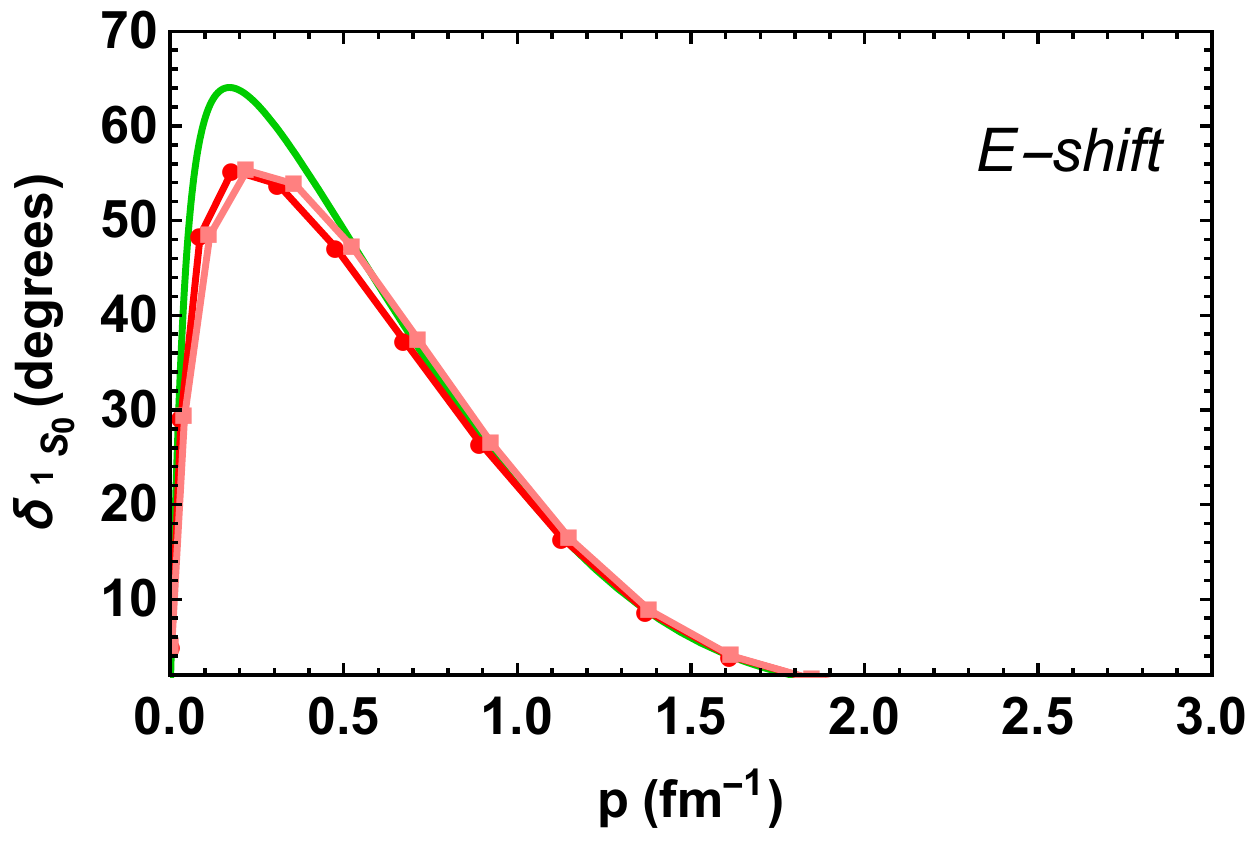}

\includegraphics[scale=0.45]{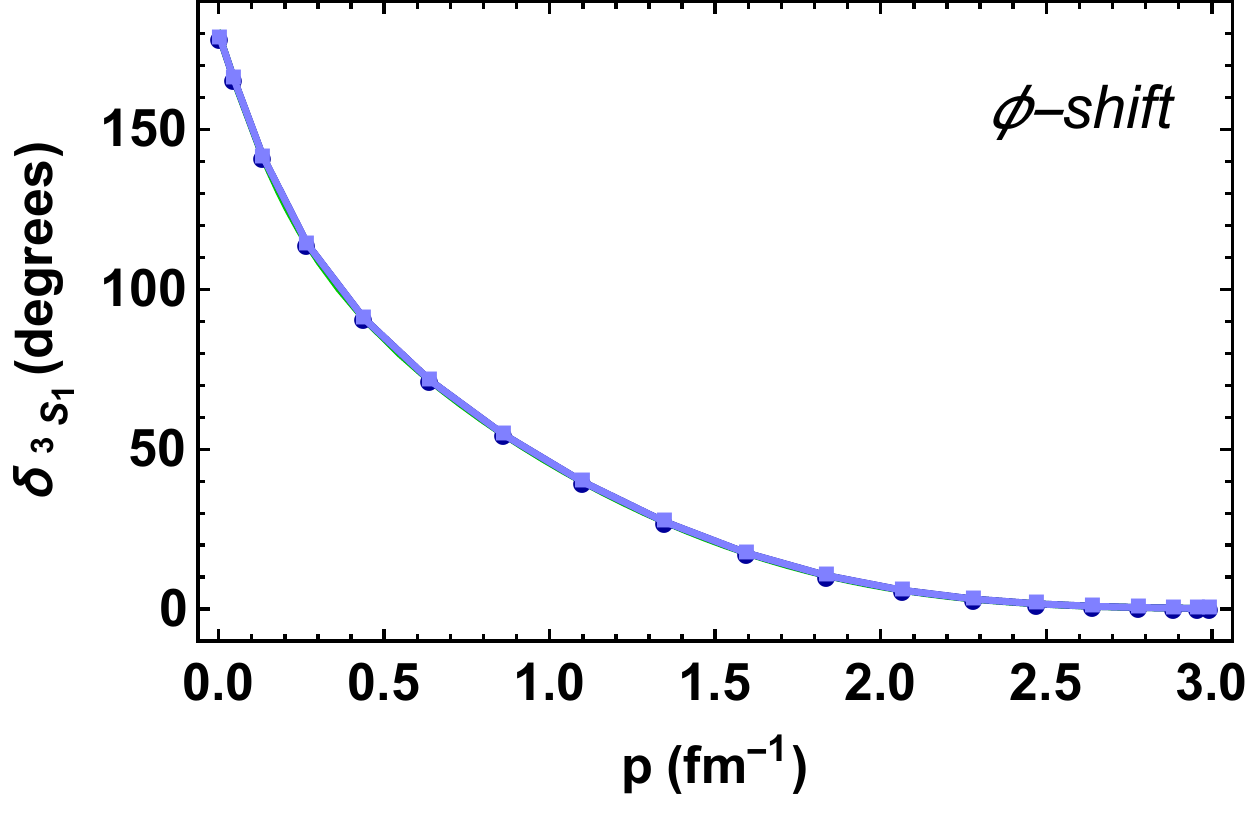}
\includegraphics[scale=0.46]{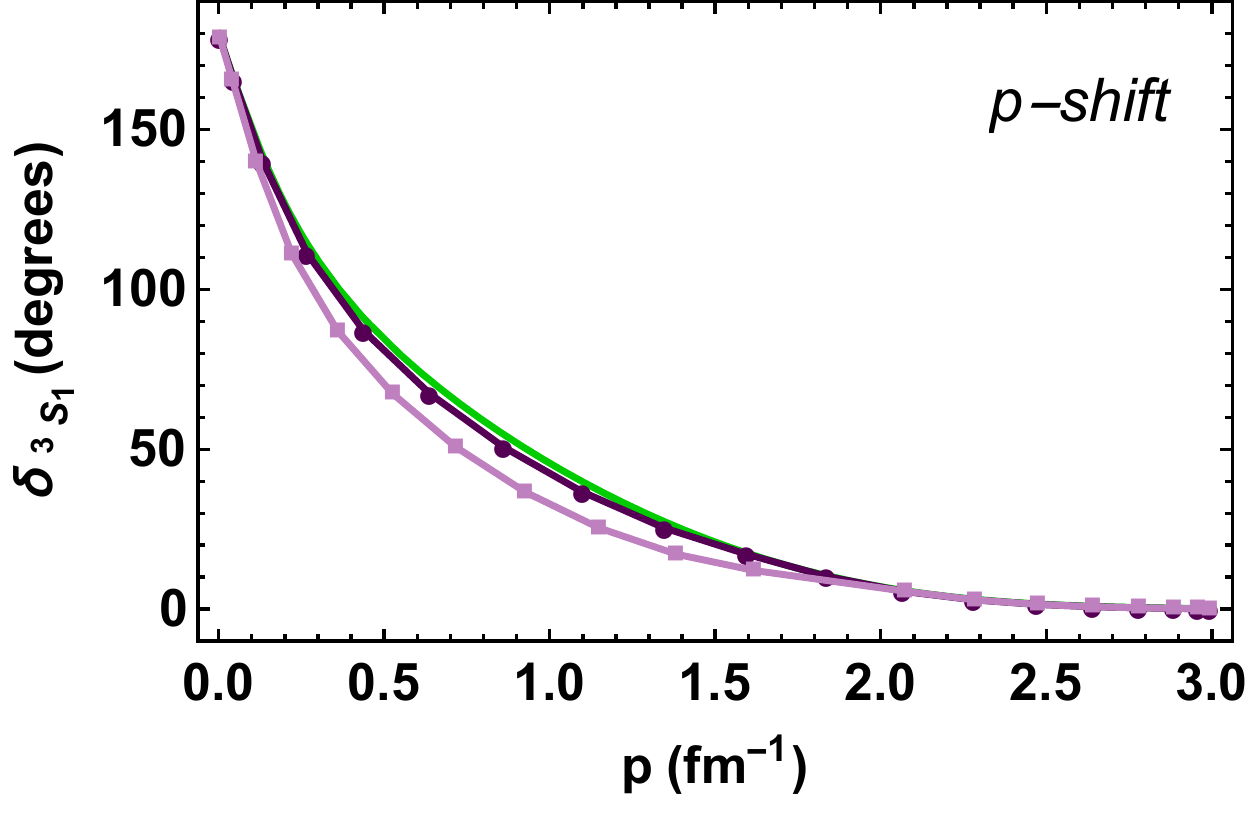}
\includegraphics[scale=0.46]{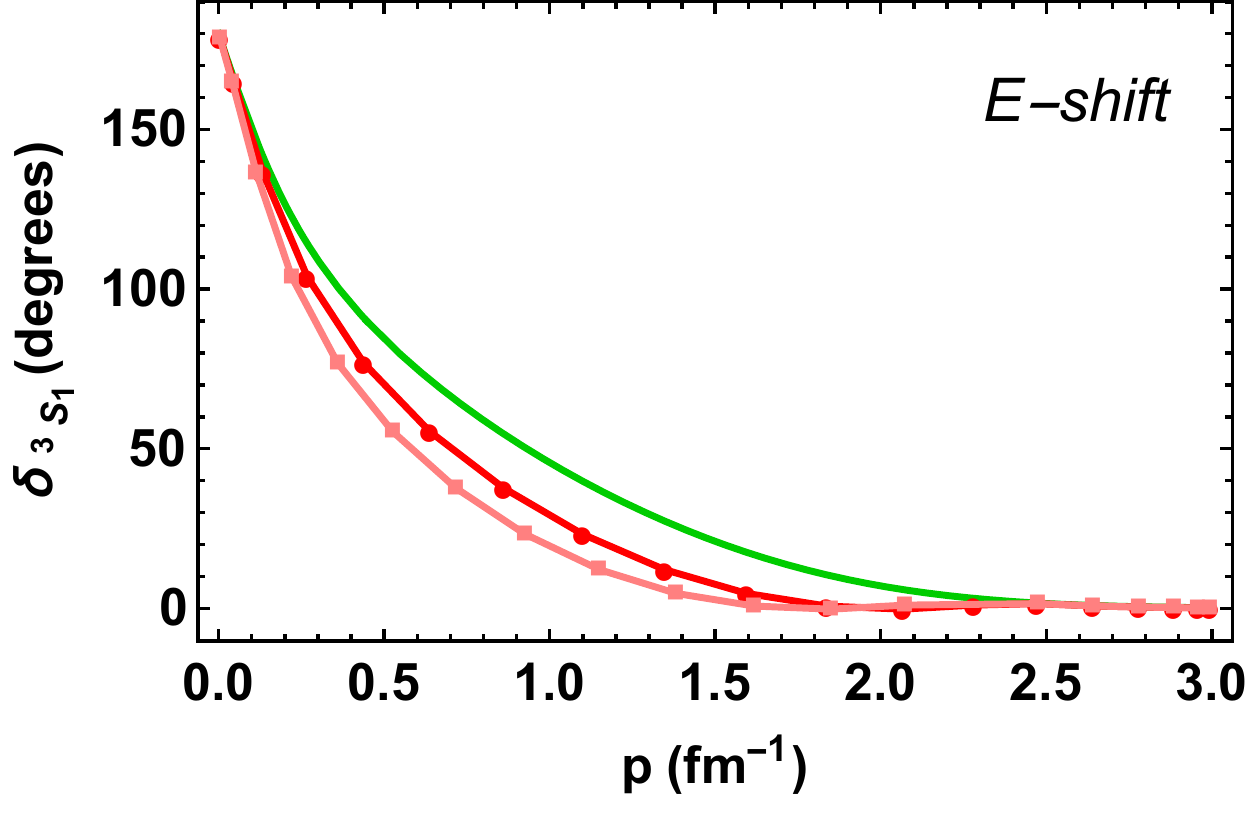}

\caption{(Color online) Phase shifts calculated using different prescriptions (as specified by a label in the figure) and compared with the exact solution (green, smooth line without markers). 
In each case the phase shifts are represented as a function of the
interacting momentum (darker line with round markers), and as a
function of the free momentum (lighter line with square markers). We
have used a grid of $N=$20 points. }
\label{Fig:prescriptions}
\end{figure*}

\begin{figure*}

\includegraphics[scale=0.45]{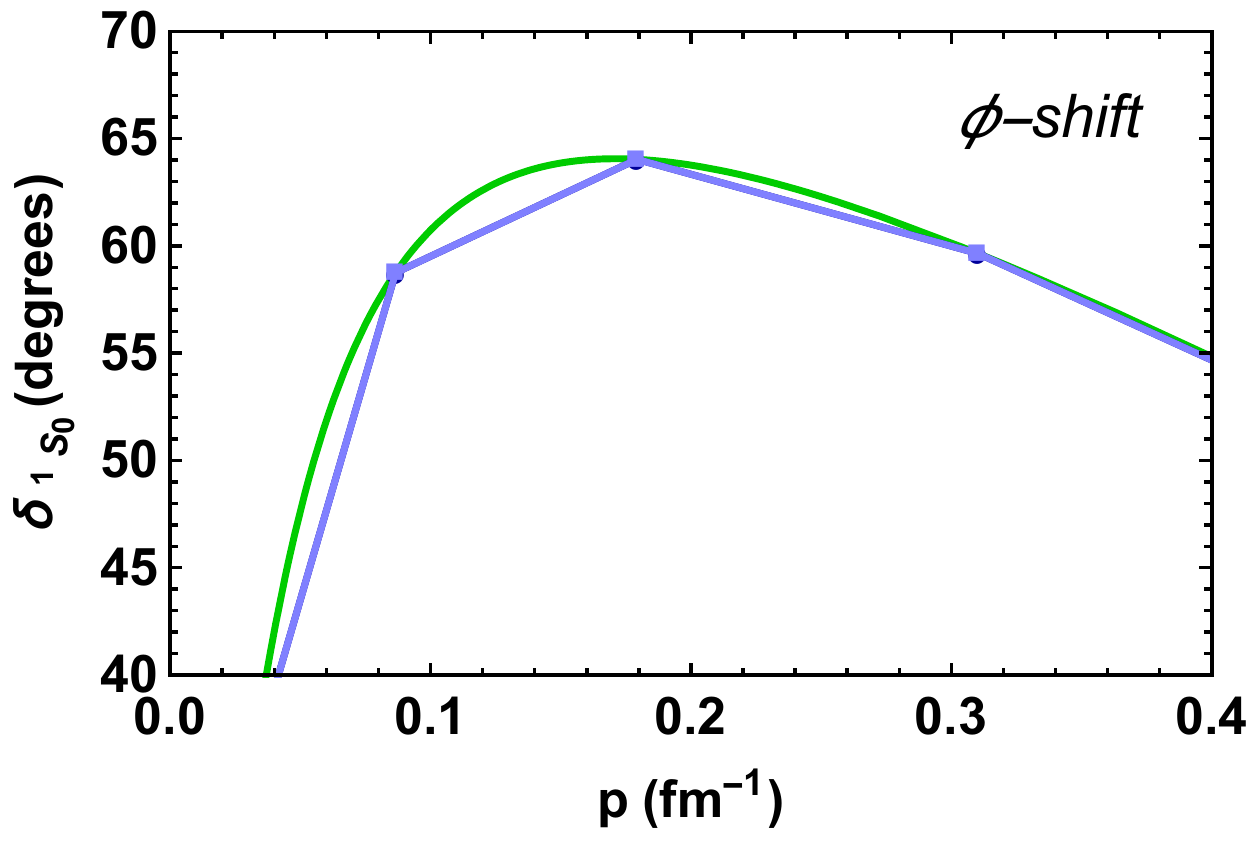}
\includegraphics[scale=0.46]{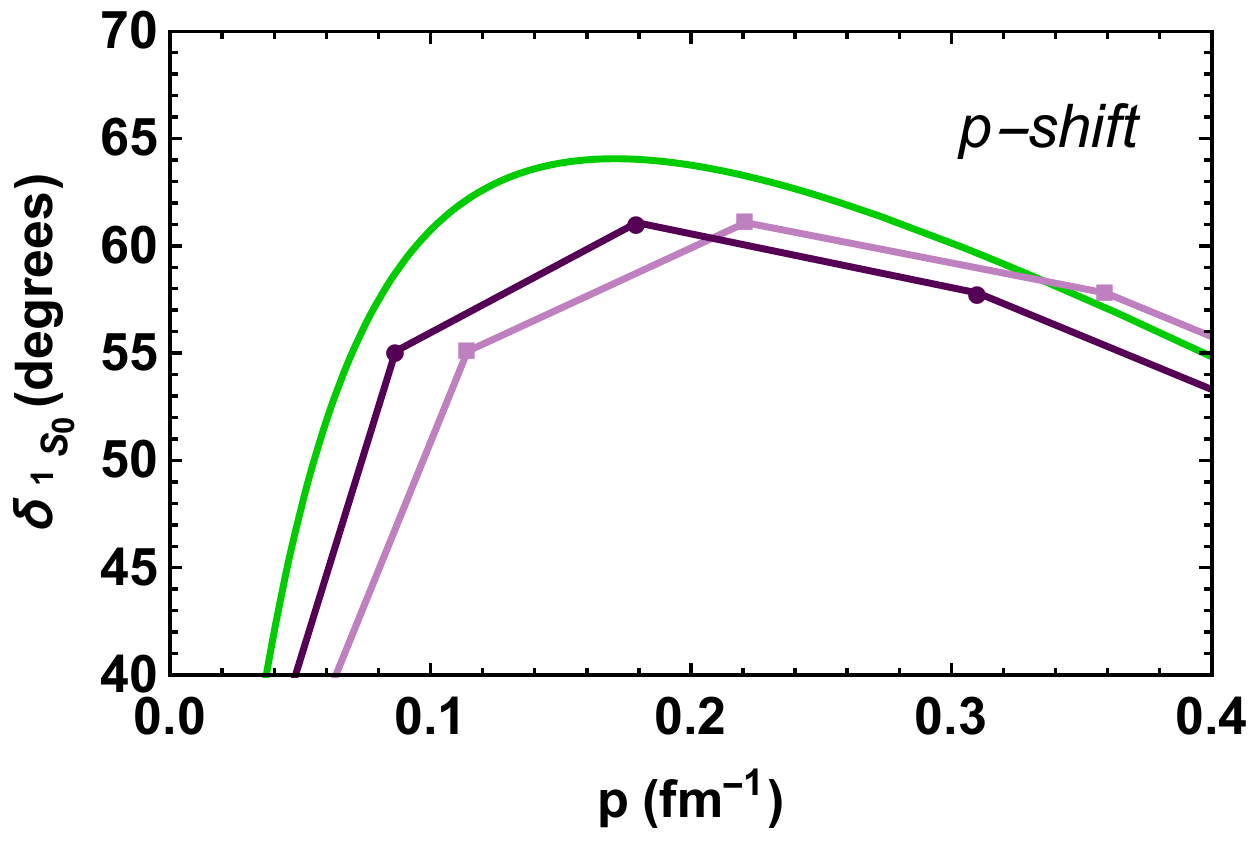}
\includegraphics[scale=0.46]{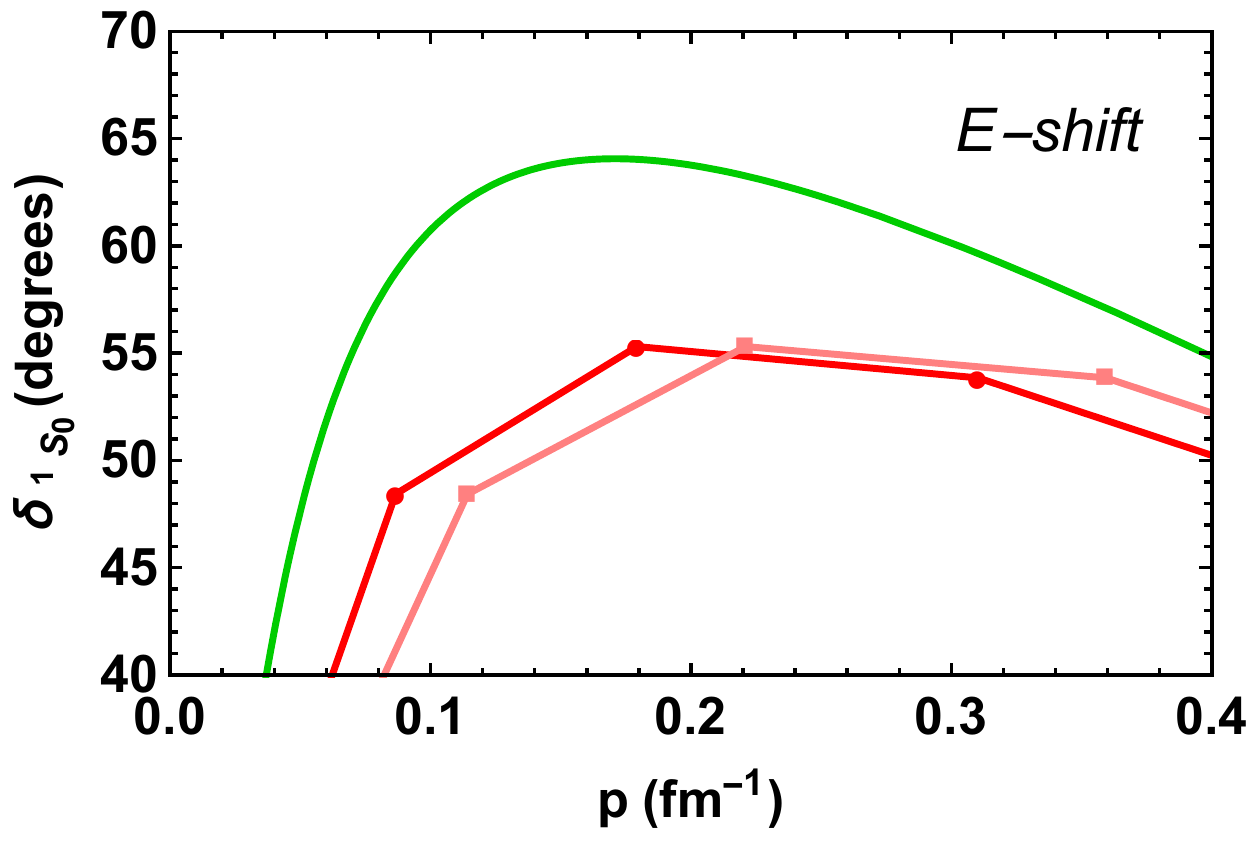}

\caption{(Color online) Selected interval of the $^1S_0$ phase shifts depicted in first row of Figure~\ref{Fig:prescriptions}, where the discrepancy between  different results is more visible.}
\label{Fig:Zoom}
\end{figure*}

\section{Numerical results}  
\label{sec:num}

We come to our numerical results and analyze how the scattering
phase-shifts obtained from the energy-shift, momentum shift and
$\phi$-shift formulas compare with the results obtained from the
standard method based on the LS equation.  Along the text, we may use
the abbreviations $p$-shift, $E$-shift, and $\phi$-shift in order to
refer to the momentum-shift, energy-shift and angle-shift
prescriptions given by Eq.~(\ref{eq:pshift}), (\ref{eq:dewitt}) and
(\ref{eq:phishift}), respectively.

We will see that the $\phi$-shift method prescription is the best and
the only one that reproduces almost exactly solution calculated in the
continuum, even for a {\it coarse} grid. In fact, as we shall show in
our numerical study, the method gives reliable predictions even for a
grid with a very small number of points. The generalization to any
momentum grid amounts to finding the variable that is uniformly
distributed along the momentum grid.

\subsection{Separable models}

In our numerical comparisons we use a separable model potential with the structure
\begin{eqnarray}
v_l(p',p)= C_l g_l(p') g_l(p) \ ,
\end{eqnarray}
where $C_l$ is positive or negative for repulsive and attractive
interactions respectively. For this form  of potential the LS equation,
Eq.~(\ref{eq:LS}) is solved by the ansatz $ T_l (p',p, E)= g_l(p')
g_l(p) T_l (E) \ , $ which inserted in Eqs.~(\ref{eq:R})-(\ref{eq:Rdel}) yields
\begin{eqnarray}
p \cot \delta_l(p) = - \frac{1}{ V_l(p,p)} \left[ 1 - \frac{2}{\pi}\dashint_0^\infty dq \, q^2\, \frac{V_l(q,q)}{p^2-q^2} \right] \ .
\end{eqnarray} 
In practice we take the toy Gaussian potential $g_\alpha (p) = e^{-p^2/L_\alpha^2}$
proposed
in \cite{Arriola:2014nia,Arriola:2016fkr} for $NN$ scattering in the $^1S_0$
and $^3S_1$ channels, where $C_\alpha$ and $L_\alpha$ are given by
$(C_{^1S_0},L_{^1S_0})= (-1.92 \, \text{fm} ,1.20 \, \text{fm}^{-1})$ and $(C_{^3S_1}, L_{^3S_1})=(-2.30 \, \text{fm}, 1.55 \, \text{fm}^{-1})$
and describe $NN$ scattering at small momenta. 
Taking these
values, we may then proceed to check the phase-shift determined by the
$p$-, $E$- and $\phi$-shift formulas, which only generates them on
grid points.

\subsection{Dependence on the momentum grid and comparison with the standard method}

In this section we study numerically how our $\phi$-shift results,
calculated in a \textit{finite} momentum grid, differ from the exact
solution in the continuum (green, smooth line in all figures) and
confront these results with the standard method based on the LS
equation.

Figure~\ref{Fig:comparisons} shows the phase shifts obtained by
solving the LS equation, Eq.~(\ref{eq:ps-LS}) (in orange), and by the
$\phi$-shift prescription Eq.~(\ref{eq:phishift}) (blue dots). Both of
them are compared with the numerical fit (smooth, green line) which
represents the exact solution. Each column of
Fig.~\ref{Fig:comparisons} corresponds to the same calculation, but
using a different number of grid points, $N=$10, 20, 50, respectively.
It is impressive how the $\phi$-shift formula reproduces exactly the
exact solution in all cases, even in the grid with least points,
$N=$10.  Conversely, we observe how the LS method converges to the
continuum as the number of points increases and a comparatively larger
number of grid points is required to reproduce the exact solution.
The LS results are worse in the case of the $^1S_0$ wave, where the
phase shifts undergo larger changes is a relatively small interval of
momentum and thus the number of points describing the curve becomes
more important.

\subsection{Comparison with different spectral-shift prescritpions}

In this section we compare the $\phi$-shift prescription with the
$p$-shift and $E$-shift ones.  Figure~\ref{Fig:prescriptions} shows,
for both channels, the resulting phase shifts obtained from the
different formulas, as indicated by a label in the corner. All of them
are compared with the exact solution.

In all cases, we may represent the results as a function of the
interacting momentum $P_n$, or as a function of the free momentum
$p_n$. Every graphic in Figure~\ref{Fig:prescriptions} shows both
curves and Figure~\ref{Fig:Zoom} shows a selected interval of the
$^1S_0$ channel, where the difference between lines is more
visible. Here $\delta (P_n)$ is represented by a darker line with
round markers, while $\delta (p_n)$ is represented by a lighter line
with square markers.  Of course, the phase shifts have the same values
but are horizontally displaced from each other by the momentum shift,
as it can be observed in the figures.  As it turns out, the
formulation as a function of the interacting momentum lies closer to
the exact solution in all cases.  This appears reasonable if one notes
that in the case of the $p$-shift formula, the phase shift must be a
function of the interacting momentum by construction.  Although there
is nothing in DeWitt's argument~\cite{DeWitt:1956be} that suggests
that the distorted momentum should be used as the independent
variable, we note that in the relativistic case and for very light
masses the $E$-shift prescription converges to the $p$-shift
one\footnote{In order to check this, one can replace the energy by the
relativistic formula $E=\sqrt{p^2+m^2}$ in Eq.~(\ref{eq:DeltaE}), and
take the limit $m\to \infty$.}.  This suggests to consider the
interacting momentum as the independent variable in all cases.

It is remarkable that in the $\phi$-shift case (first column of
Figures~\ref{Fig:prescriptions} and ~\ref{Fig:Zoom}, in blue), both
lines, the one depicted as a function of $p_n$ and the one as a
function of $P_n$, precisely overlap the exact solution and there is
no difference among both criteria. Indeed, one can barely see the
green line in this case, and the darker blue line is totally covered
by the lighter-blue line.

The fact that the $E$-shift or the $p$-shift prescriptions produce
worse results than the $\phi$-shift one is reasonable. These
prescriptions assume equal-distance separation of energy levels and
momentum levels, respectively, while in our Gauss-Chebyshev grid, this
separation occurs in the Chebyshev angle. The $E$-shift and $p$-shift
prescriptions appear thus to be a (non exact) but approximate
formula. Nevertheless, both of them still turn out to be a
considerably good approximation, since the quality of the results for
the studied $N=20$ case are by far better than those obtained through
the standard LS equation.

\section{Conclusions}
\label{sec:concl}

In this letter we have presented a new method which solves the
scattering problem by diagonalizing the corresponding Hamiltonian in a
momentum grid. This guarantees that the phase-shifts are invariant under
unitary transformations on the finite grid, unlike the usual
solutions based on the Lippmann-Schwinger equation. 

We presented and studied the predictive power of the momentum-shift
and energy-shift method for calculating phase shifts. We have proposed
a new prescription based on an argument that holds for almost any
momentum grid. The prescription requires to find the variable that
holds an equidistant space between points along the momentum grid. In
our case, the chosen grid is a Gauss-Chebyshev quadrature and the
equal spacing occurs in the angle $\phi={{\pi\over N}(n-1/2)}$. Having
identified this quantity, we follow DeWitt's
reasoning~\cite{DeWitt:1956be} and calculate phase-shift from the
$\phi$-shift produced by the interaction.  We observe that this
prescription yields remarkable good results, even for a grid with a small number of points. We have observed, furthermore,
that, in contrast to what is formulated by DeWitt, it is better to
consider the phase shifts as a function of the distorted momentum --
following the momentum-shift prescription~\cite{Reifman:1955ca} --,
instead of as a function of the free momentum.

Theses result suggest that the new $\phi$-shift method offers a
reliable tool for numerical calculations of phase shifts in a
discretized momentum grid which turns out to surpass in precision
other conventional formulas. Extensions to coupled channels and
relativistic systems are straightforward and will be analyzed in
detail elsewhere.

\section*{Acknowledgment}
This work is supported by the Spanish MINECO and European FEDER funds
(grants FIS2014-59386-P and FIS2017-85053-C2-1-P) and Junta de
Andaluc\'{\i}a (grant FQM-225).  M.G.R has been supported in part by
the European Commission under the Marie Skłodowska-Curie Action
Co-fund 2016 EU project 754446 – Athenea3i and by the SpanishMINECO’s
Juan de la Cierva-Incorporación programme, Grant Agreement
No. IJCI-2017-31531.


\end{document}